\documentclass[aps,prd,twocolumn,showpacs]{revtex4}

\usepackage{graphicx}
\usepackage{amsmath}
\usepackage{amssymb}
\usepackage{color}

\newcommand{\be}{\begin{equation}}
\newcommand{\ee}{\end{equation}}

\begin{document}

\title{Anomalous thermospin effect in the low-buckled Dirac materials}
\author{V.P.~Gusynin}
\affiliation{Bogolyubov Institute for Theoretical Physics, National Academy of Science of
Ukraine, 14-b Metrologicheskaya Street, Kiev, 03680, Ukraine}
\author{S.G.~Sharapov}
\affiliation{Bogolyubov Institute for Theoretical Physics, National Academy of Science of
Ukraine, 14-b Metrologicheskaya Street, Kiev, 03680, Ukraine}
\author{A.A.~Varlamov}
\affiliation{CNR-SPIN, University ``Tor Vergata'', Viale del Politecnico 1, I-00133 Rome,
Italy}
\date{\today }

\begin{abstract}
A strong spin  Nernst effect with nontrivial dependences on the carrier concentration and electric field applied
is expected in silicene and other low-buckled Dirac materials.
These Dirac materials can be considered as being made of two independent electron subsystems of the two-component gapped Dirac fermions.
For each subsystem the gap breaks a time-reversal symmetry and thus  plays a role of an effective
magnetic field. Accordingly, the standard Kubo formalism has to be altered by including the
effective magnetization in order to satisfy the third law of thermodynamics.
We explicitly demonstrate this by calculating the magnetization and showing how the correct thermoelectric coefficient emerges.
\end{abstract}

\maketitle

%\email{sharapov@bitp.kiev.ua}

%\affiliation{Mediterranean Institute for Fundamental Physics, Rome, Italy}

%81.05.ue 	Graphene (for structure of graphene, see 61.48.Gh; for phonons in graphene, see 63.22.Rc; for thermal properties, see %65.80.Ck; for graphene films, see 68.65.Pq; for electronic transport, see 72.80.Vp; for electronic structure, see 73.22.Pr; for %optical properties, see 78.67.Wj)

%65.80.Ck 	Thermal properties of graphene

% 72.80.Vp 	Electronic transport in graphene

% 72.25.Dc 	Spin polarized transport in semiconductors

% 73.20.At 	Surface states, band structure, electron density of states

%\keywords{GRAPHENE}

\section{Introduction}
The thermoelectric and thermomagnetic phenomena  discovered in the 19th century
turned out unexpectedly to be in the spotlight at the beginning of the 21st century. First of all, from
a practical point of view,  control of the heat fluxes and minimization of
the related losses are important factors for designing modern elements of
nanoelecronics. At the same time, the discovery of a Nernst-Ettinghausen (NE)
signal 100 times larger than its normal value in the pseudogap phase of a
high-temperature superconductor, La$_{2-x}$Sr$_{x}$CuO$_{4}$ \cite{Ong00},
followed by a similar finding (with  10$^{3}$ enhancement in magnitude)
in the fluctuating regime of conventional superconductor, Nb$_{0.15}$Si$%
_{0.85}$ \cite{Aub06}, was indicated on a sensitive and powerful tool for the study
of the microscopic properties of novel systems. In graphene where the Dirac
point can be crossed by tuning the position of the chemical potential $\mu $,
 Seebeck and Nernst effects of $\sim
50-100\,\mu V/K$ at room temperature were observed \cite%
{Wei2009PRL,Zuev2009PRL,Checkelsky2009PRB,Wang2011PRB},
which are huge compared to the nonmagnetic metals.
These measurements provided  unique information on the details of the electronic structure of the
ambipolar nature of graphene, and they correspond with the electrical
transport studies. Moreover, Seebeck and NE effects can be further enhanced
and controlled by opening a gap in the quasiparticle spectrum of the Dirac
materials \cite{Gusynin2006PRB,Sharapov2012PRB,Wang2011PRL}.

In this context, the synthesis of silicene
\cite{Lalmi2010APL,Padova2010APL,Padova2011APL,Vogt20102PRL,Lin2012APExp,Fleurence2012PRL,Chen2012PRL,Majzik2013JPCM},
a monolayer of silicon atoms forming a two-dimensional low-buckled
lattice, boosted theoretical studies of a wide class of new Dirac materials.
The honeycomb lattice of silicene can be described as in graphene in terms
of two triangular sublattices. However, a larger ionic size of silicon atoms
results in buckling of the 2D lattice. Accordingly, the sites on the two
sublattices are vertically separated at a distance $2 d \approx 0.46 %
\mbox{\AA} $. Consequently, silicene is expected \cite%
{Cahangirov2009PRL,Drummond2012PRB,Liu2011PRL,Liu2011PRB} to have a strong
intrinsic spin-orbit interaction that results in a sizable spin-orbit gap, $%
\Delta _{\text{SO}}$, in the quasiparticle spectrum opened at the Dirac
points. Moreover, by applying the electric field $%
E_z$ perpendicular to the plane, it is possible to create the on-site potential difference between the
two sublattices opening the additional gap, $\Delta_z = E_z d$, in the
quasiparticle spectrum. Similar structure and properties are also expected
in 2D sheets of Ge, Sn, P and Pb atoms \cite{Tsai2013NatComm,Xu2013PRL} (the
first three materials are coined as germanene, stanene and phosphorene).

Due to nonzero spin-orbit gap $\Delta _{\text{SO}}$ the quantum spin Hall
(QSH) effect \cite{Kane2005PRL} becomes experimentally accessible in
silicene. The latter is fundamentally related to the anomalous Hall effect in
ferromagnets \cite{Nagaosa2010RMP}. The analogy between thermomagnetic
phenomena in graphene and spintronics of the low-buckled Dirac materials
shows that the latter should also be very promising for the investigation
of the spin caloritronics phenomena \cite{Bauer2012NatMat}. Among these,
there is a particular  interest in the off-diagonal spin Nernst (SN) effect, an
analogue of the NE effect in a normal conductor subjected to a magnetic field.
It is the spin-orbit gap $\Delta _{\text{SO}}$ that plays the role of an
effective magnetic field that generates the SN effect even in the absence of
a real magnetic field in the new Dirac materials. In this paper, we will
show that due to the large value of $\Delta_{\text{SO}}\sim 10\,\mbox{meV},
$ the off-diagonal thermospin coefficient, $\beta _{xy}^{S_{z}}$, is indeed
expected to be huge in these materials while at the same time being nontrivially
dependent on the chemical potential $\mu $ and electric field $E_{z}$.

The paper is organized as follows. We begin by presenting
in Sec.~\ref{sec:models} the  model describing silicene and the basic model of the gapped two-component Dirac fermions.
The relationship between thermospin transport coefficients for silicene and thermolectric
coefficients for the basic model is considered. The qualitative analysis of the thermoelectric coefficient $\beta_{xy}$
is given using the Mott relation in Sec.~\ref{sec:qualitative}.
The specifics of the off-diagonal thermal transport and the necessity to consider magnetization currents
are discussed in Sec.~\ref{sec:Kubo-modified}.  The  magnetization of the gapped Dirac fermions is
considered in Sec.~\ref{sec:Magnetization} (the derivation is given in Appendix~\ref{sec:Appendix-magnetization}).
The final results for two-component Dirac fermions and silicene are presented in Sec.~\ref{sec:Results}.
The main results are summarized in Sec.~\ref{sec:disc}, where also the possibility of the
experimental observation of the SN in silicene is discussed.

\section{Models}
\label{sec:models}

\subsection{Model of silicene}

The low-energy physics of silicene is described by
the Hamiltonian density \cite{Liu2011PRL,Liu2011PRB,Drummond2012PRB}
\begin{equation}
\begin{split}
\mathcal{H}_{\xi }=& \sigma _{0}\otimes \lbrack \hbar v_{F}(\xi k_{x}\tau
_{1}+k_{y}\tau _{2})+\Delta _{z}\tau _{3}-\mu \tau _{0}] \\
& -\xi \Delta _{\mathrm{SO}}\sigma _{3}\otimes \tau _{3},
\end{split}
\label{Hamiltonian-density}
\end{equation}%
where the Pauli matrices $\pmb{\tau}$ and $\pmb{\sigma}$, and the unit matrices, $\tau_0$ and $\sigma_0$
act in the sublattice and spin spaces, respectively, and the wavevector $\mathbf{k}$
is measured from the $\mathbf{K}_{\xi }$ points (valleys) with $\xi =\pm $. Here we
neglected the small Rashba interaction \cite{Ezawa2012NJP}.
The Hamiltonian (%
\ref{Hamiltonian-density}), describes four kinds (two identical pairs) of
the noninteracting massive (gapped) Dirac quasiparticles with the masses $\Delta
_{\xi \sigma }/v_{F}^{2}$, where $v_{F}$ is the Fermi velocity, $\sigma
(=\pm )$ is the spin, and the valley-dependent gap $%\begin{equation}
%\label{gap-valley-spin}
\Delta _{\xi \sigma }=\Delta _{z}-\xi \sigma \Delta _{\text{SO}}.%
%\end{equation}
$

The QSH effect in silicene occurs due to the presence of two subsystems with $%
\sigma = \pm$ each exhibiting the quantum Hall effect. The corresponding chiral
edge states are spin-polarized and form a time-reversed pair to recover the
overall time-reversal symmetry (the Kane-Mele scenario \cite{Kane2005PRL}).
The spin Hall conductivity can be expressed \cite%
{Sinitsyn2006PRL,Dyrdal2012PSS} in terms of the electric Hall conductivity $%
\sigma_{xy} (\Delta)$ for the two-component Dirac fermions with the gap $\Delta$
(see the Hamiltonian (\ref{Dirac-Hamiltonian}) below) by the relation
\begin{equation}  \label{Hall2spin-Hall}
\sigma_{xy}^{S_z} = - \frac{\hbar}{2e} \sum_{\xi, \sigma = \pm } \xi \sigma
\sigma_{xy} (\Delta \to \Delta_{\xi \sigma}).
\end{equation}
The factor $-\hbar/2 e$ indicates that in the off-diagonal correlation
function of the two electric currents, one electric current is replaced by
the spin current.

Being subjected to the temperature gradient $\pmb{\nabla} T$, the spin-polarized chiral
edge states loose their time-reversal symmetry and the spin current  $\mathbf{j}^{s}$  flows.
The latter is related to $\pmb{\nabla} T$
by means of the thermospin tensor, $\hat{\beta}^{s}$, via
\cite{Bauer2012NatMat} $\mathbf{j}^{s}=-\hat{\beta}^{s}\pmb{\nabla}T$.
Analogously to the spin Hall conductivity, the off-diagonal component $\beta
_{xy}^{S_{z}}$ can be obtained from (\ref{Hall2spin-Hall}) by the substitution $%
\sigma _{xy}^{S_{z}}\rightarrow \beta _{xy}^{S_{z}}$ and $\sigma
_{xy}(\Delta )\rightarrow \beta _{xy}(\Delta )$, where $\beta _{xy}(\Delta )$
is the standard thermoelectric coefficient for the two-component Dirac
fermions. Thus in the absence of valley mixing, the study of the spin
transport coefficient is reduced to an investigation of the electric
transport for the two-component gapped Dirac fermions.

\subsection{Model of two-component Dirac fermions}

The corresponding Hamiltonian density is
\begin{equation}
\mathcal{H}=\hbar v_{F}(k_{x}\tau _{1}+k_{y}\tau _{2})+\Delta \tau _{3}-\mu
\tau _{0}.  \label{Dirac-Hamiltonian}
\end{equation}%
This model with broken  time-reversal symmetry
provides a simple realization of the anomalous Hall and thermoelectric effects.
Its main merit is the possibility of obtaining  simple approximate analytical expressions in
the presence of spin-independent random potential with
Gaussian correlations \cite{Sinitsyn2006PRL,Sinitsyn2007PRB,Nagaosa2010RMP}.
The two-component fermion model (\ref{Dirac-Hamiltonian}) is considered in
Secs.~\ref{sec:qualitative}, \ref{sec:Kubo-modified} and \ref{sec:Magnetization}.

\section{Qualitative analysis}
\label{sec:qualitative}

A qualitative evaluation of the thermoelectric coefficient $\beta
_{xy}(\Delta )$ can be obtained basing on the Mott relation,
\begin{equation}
\label{Mott-formula}
\beta _{xy}=-\frac{\pi ^{2}k_{B}^{2}}{3e}T
\frac{\partial \sigma _{xy}(\mu,\Delta ,T=0)}{\partial \mu },
\end{equation}%
where $k_{B}$ is the Boltzmann constant.
In the clean limit at $T=0$ one finds \cite{Sinitsyn2006PRL,Sinitsyn2007PRB}
\begin{equation}
\label{Hall-clean-final}
\sigma _{xy}=-\frac{e^{2} \mathrm{sgn}\, (\Delta) }{4\pi \hbar }%
\begin{cases}
1, & |\mu |\leq |\Delta |, \\
|\Delta| /|\mu |, & |\mu |>|\Delta |,%
\end{cases}
\end{equation}%
from which we can draw two conclusions:
\textit{i}) For $|\mu |>|\Delta |$, we obtain that $\beta
_{xy}=-(k_{B}/e)(\pi e^{2}/12\hbar )(\Delta \mbox{sgn}\,(\mu) /\mu ^{2})k_{B}T$.
Then the Nernst signal  $e_{y}(T)\equiv -E_{y}/ \nabla _{x} T$, where  $E_{y}$
is the electric field in $y$-direction, can be estimated as
\vspace{-1mm}
\begin{equation}
e_{y}(T) \approx \frac{\beta _{xy}}{\sigma
_{xx}}=-\left( \frac{k_{B}}{e} \right) \frac{\pi e^{2}}{12\hbar \sigma _{xx}}
\frac{ k_{B}T \Delta \mbox{sgn}\,(\mu) }{\mu ^{2}}.
\end{equation}
%\vspace{-1mm}
Here we assumed
that the diagonal conductivity is much larger than the Hall conductivity, $\sigma _{xx}\gg |\sigma _{xy}|$.

The main feature of the Dirac materials is
that the value of the chemical potential $\mu $  can be tuned
as close as possible to the regime with $e_{y}(T)\sim
k_{B}/e\sim 86\,\mu V/K$. This is exactly how one gains from 3 to 4 orders of
magnitude in the Nernst signal as compared to the normal nonmagnetic metals,
where $e_{y}$ is negligibly small ($\sim 10\,nV/K$ per Tesla).\newline
\textit{ii}) Our simple estimate also shows that  Mott's formula fails
near $|\mu |=|\Delta |$ when the conductivity $\sigma _{xy}(\mu ,\Delta ,T=0)$
changes abruptly. Indeed, as discussed recently in \cite{Sharapov2012PRB} (see also references therein), when the gap is
present in the quasiparticle spectrum, one should use the microscopic
approach. The same is true for the SN effect:
as of yet it has been studied mostly
using a formula analogous to  Mott's formula written for the spin conductivity
\cite{Bao2012ChinPhys,Dyrdal2012JPCM,Rothe2012PRB}
(see also the numerical study \cite{Liu2010SSC} based on the Landauer-Buttiker formula).
%Nevertheless, the Mott's formula is used very widely in the study of the NE
%effect in view of its simplicity and possibility to avoid
%the difficulties that we will highlight below.

\section{Modified Kubo formula}
\label{sec:Kubo-modified}

The study of the off-diagonal thermal transport
in the framework of the Kubo formalism is a delicate issue.
It was  firstly understood 50 years ago by Obraztsov \cite{Obraztsov1964FTT} that, in conjunction
with the Kubo-like response on the temperature gradient,
 magnetization currents must be taken into account in order to satisfy the Onsager principle of
the symmetry of the kinetic coefficients. It is worthwhile to mention that
this problem has been readdressed in almost every decade \cite%
{Streda1977JPC,Jonson1984PRB,Oji1985PRB,Cooper1997PRB,Xiao2006PRL} due to its importance
for the quantum Hall effect, NE in fluctuating superconductors, etc. In the
problem under consideration the account for magnetization currents turns out to
be crucial not only in order to get the correct coefficient in $\beta_{xy}$
for the two-component gapped Dirac fermions, but first and foremost for the validity of the third law of thermodynamics.

We will show below, that the mere calculation of
\begin{equation}
\label{Kubo-def}
\tilde{\beta}_{xy}=-\frac{\hbar}{T} \lim_{\omega \rightarrow 0} \frac{Q_{xy}^{eq\left(
R\right) }(\omega )}{\omega}
\end{equation}
(here $Q_{xy}^{eq\left( R\right)}$ is the retarded
response function of electric and heat currents) in the Kubo formalism results in the expression that in the
low-temperature limit is presented in the form of the Laurent series: $a_{-1}/T+a_0 +a_1 T+ \cdots$.
At the same time, it is clear that at zero temperature
the thermoelectric tensor must become zero: it describes the transport of entropy, which, in accordance with
the third law of thermodynamics, becomes zero when $T \to 0$.  In the presence of an effective magnetic field,
the off-diagonal thermal transport coefficient $\tilde{\beta}_{xy}$
has to be enriched \cite{Obraztsov1964FTT} by including of the magnetization $M_z$ term,
so that
%\begin{equation}
%\label{tilde2beta}
$\beta _{xy}=\tilde{\beta}_{xy}+ cM_{z}/T$ ($c$ is the velocity of light).
%\end{equation}.
The latter exactly cancels out both $a_{-1}/T$ and $a_0$ terms in the complete expression
for  $\beta_{xy}$,  making it proportional to the absolute temperature in the vicinity of $T=0$ and
reconciling the theory with the basic principles of thermodynamics.

The  above-mentioned electric-heat currents linear response function in the
Matsubara representation can be presented as the bubble of two Green's
functions (GF)%
\begin{equation}
\label{Q-with-vertex}
\begin{split}
& Q_{\alpha \beta }^{eq}\left( \Omega _{m}\right) =\sum_{\epsilon_n}  \int \frac{d^2 k}{(2 \pi)^2}
\mbox{tr}%
\left[ \pmb{\Upsilon}_{\alpha }^{\left( e\right) }(\epsilon_n +\Omega_m,\epsilon_n ) \right. \\
& \left. \times G(\epsilon_n +\Omega _{m},\mathbf{k})
\Upsilon_{\beta}^{\left( q\right) }(\epsilon_n, \epsilon_n +\Omega _{m})G(\epsilon_n ,\mathbf{k}) \right].
\end{split}
\end{equation}%
Here $G(\epsilon _{n},\mathbf{k})=\left[ i\epsilon _{n}\tau _{0}-\mathcal{H}%
-\Sigma (i\epsilon _{n})\right] ^{-1}$ is the two-component gapped Dirac
fermions' GF with the self-energy
$\Sigma^{R}(\epsilon )=-i\Gamma_{0}(\epsilon )\tau _{0}-i\Gamma _{1}(\epsilon )\tau _{3},$
$\Upsilon_{\beta }^{\left( q\right) }(\epsilon_n, \epsilon_n +\Omega _{m})=i\left(
\epsilon _{n}+\Omega _{m}/2\right) v_{F}\tau _{\beta }$ is the heat current
vertex for non-interacting fermions, and finally,
$\pmb{\Upsilon }_{\alpha }^{\left( e\right) }(\epsilon_n +\Omega _{k},\epsilon_n )$ is the electric current vertex
renormalized by the impurity scattering. $\Gamma_{0}(\epsilon )$ and $\Gamma_{1}(\epsilon )$
are the corresponding scattering rates.

To see the role of magnetization it is very instructive to start our analysis from the clean limit, $%
\Sigma ^{R}(\epsilon )=-i\Gamma _{0}(\epsilon )\tau _{0}$ with $\Gamma _{0}(\epsilon ) = \Gamma_0 \rightarrow 0$,
and the bare vertex, $\Upsilon_{\alpha }^{\left( e\right) }(\epsilon_n +\Omega_m,\epsilon_n)= - ev_{F}\tau _{\alpha}$.
In this case one can explicitly calculate the Kubo and magnetization parts
of the off-diagonal thermoelectric coefficient $\beta_{xy}$.
We obtain that the Kubo part (\ref{Kubo-def}) can be written in the form
\begin{equation}
\label{Kubo-part}
\begin{split}
\tilde{\beta}_{\alpha\beta}= &  \frac{e\hbar v_F^2}{4\pi T}\int\limits_{-\infty}^\infty d\epsilon
\left(- \frac{\partial f(\epsilon)}{\partial \epsilon} \right)\epsilon \\
&\times {\rm Tr}\left[\tau_\alpha G^R\tau_\beta (G^R -G^A)-\tau_\alpha(G^R-G^A)\tau_\beta G^A\right]\\
+ &\frac{e\hbar v_F^2}{4\pi T}\int\limits_{-\infty}^\infty d\epsilon\, f(\epsilon)\epsilon
{\rm Tr}\left[\tau_\alpha \frac{dG^R}{d\epsilon}\tau_\beta G^R  \right.\\
& \left. -\tau_\alpha G^R\tau_\beta \frac{dG^R}{d\epsilon} -
\tau_\alpha \frac{dG^A}{d\epsilon}\tau_\beta G^A+\tau_\alpha G^A\tau_\beta
\frac{dG^A}{d\epsilon}\right],
\end{split}
\end{equation}
where the retarded (advanced) GF is $G^{R,A}(\epsilon,\mathbf{k}) = G(\epsilon _{n} \to \epsilon \pm i 0 ,\mathbf{k})$ and
$\mbox{Tr}$ denotes the integration over $\mathbf{k}$ and, as in (\ref{Q-with-vertex}), the trace over sublattice indices.
The Kubo contribution  (\ref{Kubo-part}) along with
the standard term containing the derivative of the Fermi distribution, $\partial f (\epsilon)/\partial \epsilon$,
also contains the term with the integral containing the Fermi distribution, $f ( \epsilon ) = 1/[\exp(\epsilon/k_B T)+1 ]$
itself.
It is the latter that in the low temperatures limit $T\rightarrow 0$
produces  the diverging  part,
\begin{equation}
\label{tilde-beta}
\tilde{\beta}_{xy}=-\frac{e}{4\pi \hbar T}[\Delta \mbox{sgn}\,(\mu) \theta
(|\mu |-|\Delta |)+\mu \mbox{sgn}\,(\Delta) \theta (|\Delta |-|\mu |)].
\end{equation}
Thus we see that even in the absence of  a real external magnetic field $B$,
the calculation of $\beta_{xy}$ in the model (\ref{Dirac-Hamiltonian}) using the Kubo formula reveals a
difficulty very similar to the problem solved by Obraztsov \cite{Obraztsov1964FTT}.
It is the gap $\Delta$ that plays the role of the external magnetic field in Eq.~(\ref{tilde-beta}).
Below we will explicitly  show that accounting for the magnetization term $cM_{z}/T$  removes the
divergence in $\beta_{xy}$.

\section{Magnetization}
\label{sec:Magnetization}

Since the Hamiltonian (\ref{Dirac-Hamiltonian}) breaks the time-reversal symmetry,
the intrinsic magnetization (magnetic moment per unit volume) \cite{Streda1977JPC,Jonson1984PRB,Oji1985PRB}
\begin{equation}
\label{M}
M_{z}=\frac{e v_F}{2 c}\int\limits_{-\infty}^\infty
d \epsilon\ f(\epsilon){\rm Tr}\left[\delta(\epsilon \tau_0 - \mathcal{H})(\hat{r}_\beta\tau_\alpha-\hat{r}_\alpha\tau_\beta)\right]
\end{equation}
is indeed expected to be nonzero.
However, an attempt to calculate
$M_z$  from the definition (\ref{M}) fails due to the difficulties that were resolved only recently
(see Ref.~\onlinecite{Chen2011PRB} and references therein). It is
the unboundedness of the coordinate operator $\hat{r}_\alpha$ with $\alpha =x,y$  that does not allow
one to derive  $M_z$ directly.

To overcome this problem, we start from the GF of the charge carriers in the magnetic field written
in the coordinate representation.
Having the GF, it is already straightforward to calculate the carrier density $\rho(\mu, T,B)$.
The thermodynamic potential $\Omega(\mu,T,B)$ can be obtained by integrating the relationship
$\rho = -\partial \Omega/\partial \mu $ over $\mu$. Finally, the magnetization is derived as
$M_z = - \partial \Omega/\partial B$
(all the details are provided in Appendix~\ref{sec:Appendix-magnetization}).
Let us stress that since the time-reversal symmetry is broken, a finite field-independent contribution to $M_z$ appears.

It follows from Eq.~(\ref{M-Gamma}) that for
$T \gg \Gamma_0$, the magnetization takes the especially simple form
\begin{equation}
\label{M-final}
\! \!
\frac{M_{z}}{T}=\frac{e\,\mathrm{sgn}\,( \Delta)}{4\pi \hbar c}\left[ \ln \cosh \frac{%
\mu +|\Delta |}{2 k_B T}-\ln \cosh \frac{\mu -|\Delta |}{2 k_B T}\right].
\end{equation}
Remarkably, in the limit $T \to 0$ the asymptotic expression of Eq.~(\ref{M-final})
reduces to Eq.~(\ref{tilde-beta})  but with the opposite sign.
This restores the validity of the third law of thermodynamics.
Since  $\beta_{xy}$ describes the transversal entropy transport it must identically become zero at $T=0$.
Equation (\ref{M-final}) illustrates in a spectacular way how  the gap $\Delta$  induces the anomalous magnetic moment.
%At $T =0$ we see that $M_z \sim \mu \, \mbox{sgn} \, \Delta$
%for $|\mu| < |\Delta|$ and  $M_z \sim \Delta \, \mbox{sgn} \, \mu$  for $|\mu| > |\Delta|$.

\section{Results}
\label{sec:Results}

Finally the off-diagonal transport coefficients
$\sigma_{xy} (\Delta)$ and $\beta_{xy} (\Delta)$ can be presented in the standard form
\begin{equation}
\label{Kubo}
\left\{
\begin{array}{c}
\sigma _{xy} \\
\beta_{xy} \\
\end{array}%
\right\}  =  \frac{e^{2}}{\hbar } \int \limits_{-\infty }^{\infty } d \epsilon
\left[ -\frac{\partial f (\epsilon)}{\partial \epsilon} \right]
 \left\{
\begin{array}{c}
-1 \\
\dfrac{\epsilon}{eT} \\
\end{array}%
\right\}  \mathcal{A}_{H}(\mu +\epsilon ,\Delta ),
\end{equation}%
where all specific information about the model and the characteristics of elastic scattering is contained in the zero temperature
Hall conductivity $\sigma_{xy}(\mu,\Delta,T=0)= -(e^2/\hbar) \mathcal{A}_{H}(\mu, \Delta)$. The analogous result was obtained by
Smr\v{c}ka and St\v{r}eda  \cite{Streda1977JPC} for
nonrelativistic fermions in a magnetic field.

\subsection{Two-component Dirac fermions}

In the clean case, in the bubble approximation
the function $\mathcal{A}_{H}$ with the level broadening acquires the form
\begin{equation}
\label{Hall-kernel}
\begin{split}
& \mathcal{A}_{H}^{\left( cl\right) }(\epsilon ,\Delta )=\frac{\Delta }{4\pi
^{2}}\left[ \frac{1}{\epsilon }\left( \arctan \frac{|\Delta |+\epsilon }{%
\Gamma _{0}}-\arctan \frac{|\Delta |-\epsilon }{\Gamma _{0}}\right) \right.
\\
& \left. +\frac{1}{|\Delta |}\left( \arctan \frac{\epsilon +|\Delta |}{%
\Gamma _{0}}-\arctan \frac{\epsilon -|\Delta |}{\Gamma _{0}}\right) \right] .
\end{split}
\end{equation}%
Accordingly, for
$T \to 0$ $\left( \text{but }T\gg \Gamma _{0}\right) $ we obtain
\begin{equation}
\label{beta-clean}
\beta _{xy}^{\left( cl\right) }=  - \beta_0  \frac{\pi k_B T}{12} \frac{\Delta \,\mbox{sgn}\,(\mu) }{\mu^2} \theta (\mu^2 - \Delta^2) ,
\end{equation}
where $\beta_0 = k_B e/\hbar$.
It is easy to see that Eq.~(\ref{beta-clean})
also directly follows from the Mott relation (\ref{Mott-formula}) and the conductivity
(\ref{Hall-clean-final}). However, the general expression (\ref{Kubo}) allows us to
investigate the vicinity of the point $|\mu| = |\Delta|$, where the Mott result (\ref{beta-clean})
fails.

The influence of disorder on the Hall conductivity of the gapped Dirac fermions was studied in \cite{Sinitsyn2006PRL,Sinitsyn2007PRB}.
The authors  found the dressed vertex $\pmb{\Upsilon }_{\alpha }^{\left( e\right) }$  in the ladder
approximation. Accordingly, in the presence of disorder the kernel $\mathcal{A}_{H}$ takes the form
\begin{equation}
\label{Hall-kernel-vertex}
\begin{split}
& \mathcal{A}_{H}^{\left( d\right) }(\epsilon ,\Delta )=\frac{\mbox{sgn}%
\,(\Delta) }{4\pi }\theta (\Delta ^{2}-\epsilon ^{2}) + \frac{\Delta }{4\pi |\epsilon |} \\\
& \times \left[ 1+\frac{4(\epsilon ^{2}-\Delta
^{2})}{\epsilon ^{2}+3\Delta ^{2}}+\frac{3(\epsilon ^{2}-\Delta ^{2})^{2}}{%
(\epsilon ^{2}+3\Delta ^{2})^{2}}\right] \theta (\epsilon ^{2}-\Delta ^{2}).
\end{split}
\end{equation}%
The important feature of $\mathcal{A}_{H}^{\left( d\right) }$ is that, in contrast to $\mathcal{A}_{H}^{\left( cl\right) }$,
it is independent of the disorder potential strength and of the impurity concentration  encoded in the scattering rates
$\Gamma_0 (\epsilon)$ and $\Gamma_1 (\epsilon)$.
Comparing the kernels $\mathcal{A}_{H}^{\left( d\right) }$ and $\mathcal{A}_{H}^{\left( cl\right) }$,
one can see that the approximation of the disorder effects by the level broadening is insufficient
even in the weak disorder limit \cite{Sinitsyn2006PRL,Sinitsyn2007PRB}, and it leads to
drastic changes in the behavior of $\sigma_{xy}$ and $\beta_{xy}$.

\begin{figure}[ht]
\includegraphics[width=4.1cm]{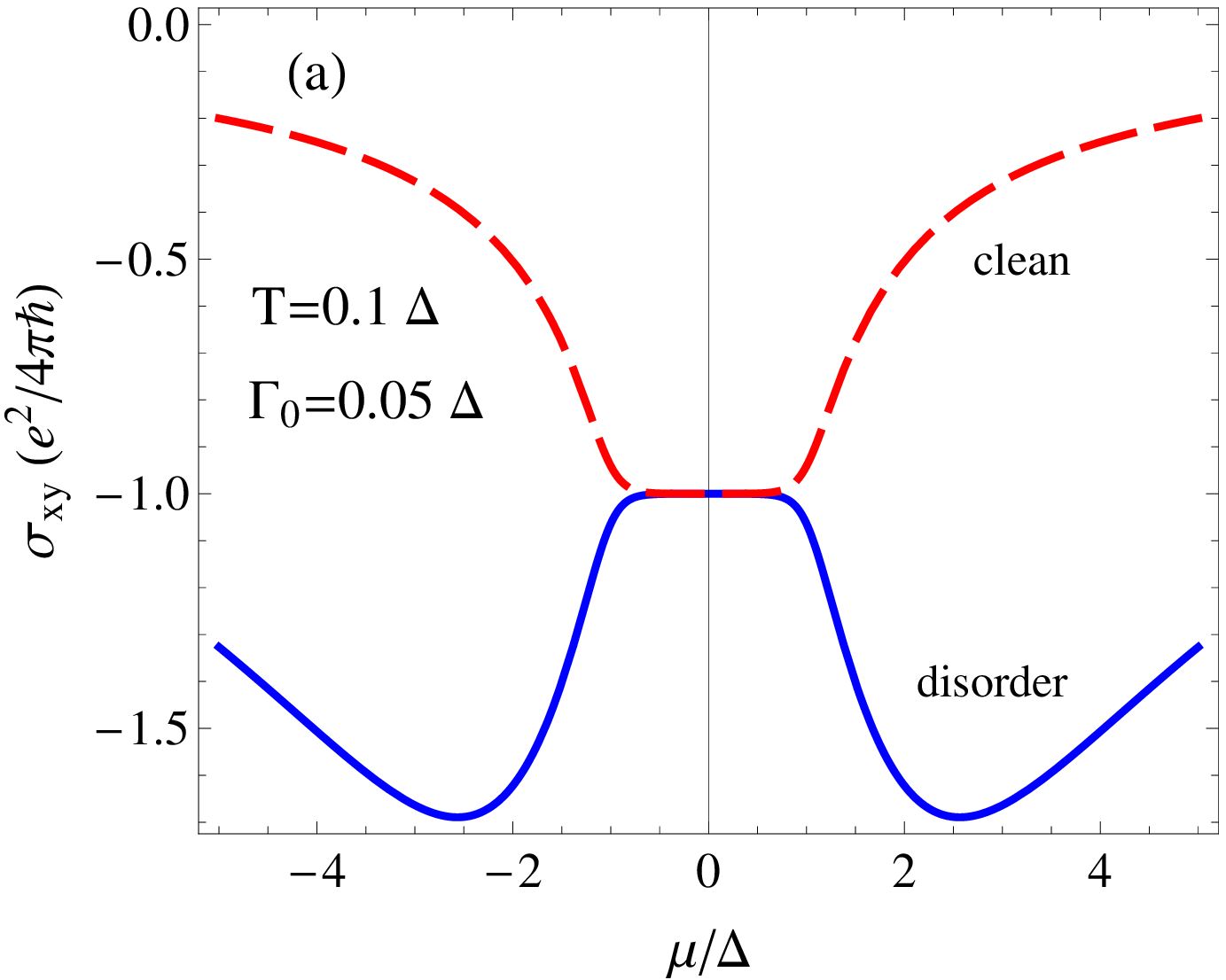}%\newline
\includegraphics[width=4.1cm]{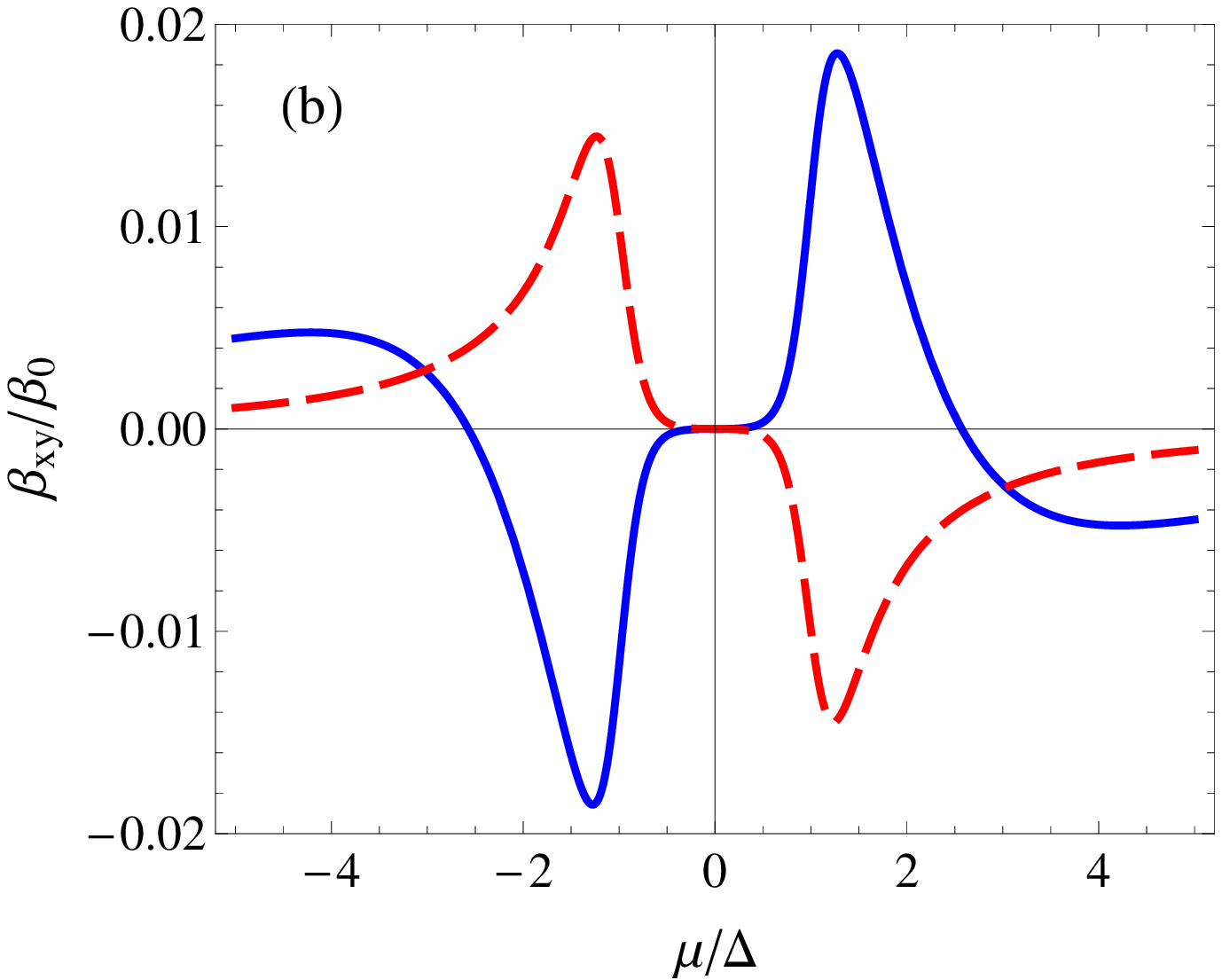}
\caption{(Color online) Left panel (a): electrical Hall conductivity $\sigma_{xy}(\mu)$
in units of  $\sigma_0 = e^2/(4 \pi \hbar)$;
right panel (b): thermoelectric coefficient $\beta_{xy}(\mu)$ in units of  $\beta_0 = k_B e/\hbar$
as functions of the chemical potential $\mu$ in the units of $\Delta>0$.
}
\label{fig:1}
\end{figure}

The dependences $\sigma_{xy}(\mu)$ and $\beta_{xy}(\mu)$
are plotted in the left and right panels of Fig.~\ref{fig:1}, respectively.
The dashed (red) and the solid (blue) curves correspond to the calculations done
using the kernels $\mathcal{A}_{H}^{\left( cl\right) }$ and $\mathcal{A}_{H}^{\left( d\right) }$, respectively.
We took $T = 0.1 \Delta$ and  $\Gamma_0 = 0.05 \Delta$.
One can see that $\sigma_{xy}$ and $\beta_{xy}$
are even and odd functions of $\mu$, respectively. On the contrary,
in the case of a real magnetic field, $\sigma_{xy}$ and $\beta_{xy}$ are odd and even functions of $\mu$, respectively.
In this respect, a positive sign of the Nernst signal near $\mu =0$ is regarded as one of the fingerprints of the
Dirac quasiparticles \cite{Checkelsky2009PRB,Gusynin2006PRB}. In the case of the anomalous Hall and Nernst effects,
the sign of $\beta_{xy}$ also remains very informative. One can see that
the presence of the disorder vertex drastically changes the pattern of the sign changes in
$\beta_{xy}(\mu)$. The nonmonotonic dependence of $\sigma_{xy}(\mu)$ on the electron ($\mu >0$) or
hole ($\mu <0$) parts of the carriers results in new nontrivial zeros of $\beta_{xy}(\mu)$.
Using the Mott relation, one finds that these zeros are
at $\mu = \pm \sqrt{3+ 2 \sqrt{3}} |\Delta| \approx \pm 2.5 |\Delta|$.

\subsection{The results for silicene}

We return now to the discussion of the silicene model (\ref{Hamiltonian-density}).
We calculate the spin Hall conductivity  $\sigma_{xy}^{S_{z}}$ from Eq.~(\ref{Hall2spin-Hall}) and
the thermospin coefficient $\beta_{xy}^{S_{z}}$ from its analog  using  Eq.~(\ref{Kubo})
for the two-component Dirac fermions.

For reference, we begin with the kernel $\mathcal{A}_{H}^{\left( cl\right) }$.
The spin Hall conductivity at $T=0$ and the zero sublattice asymmetry gap $\Delta_z =0$   directly follows
from Eq.~(\ref{Hall-clean-final}) and reads \cite{Kane2005PRL,Sinitsyn2006PRL}
\begin{equation}
\label{SH-ref}
\begin{split}
\sigma_{xy}^{S_z} = &- \frac{e}{2\pi}\, \mbox{sgn} \, (\Delta _{\text{SO}}) \\
& \times
\left[\theta(|\Delta _{\text{SO}}| - |\mu|)
+ \frac{|\Delta _{\text{SO}}|}{|\mu|} \theta(|\mu| - |\Delta _{\text{SO}}|) \right] .
\end{split}
\end{equation}
Let us note that although the spin-orbit gap $\Delta _{\text{SO}}$ does not break the time-reversal symmetry, one
can check that the Kubo contribution for $\Delta_z =0$ and $T \to 0$
is
\begin{equation}
\tilde{\beta}_{xy}^{S_z}= \frac{e}{\pi \hbar T}\Delta _{\text{SO}} \mbox{sgn}\,(\mu) , \qquad |\mu| > |\Delta _{\text{SO}}|.
\end{equation}
This divergence, as above, is compensated by the ``spin magnetization''
\begin{equation}  \label{M2spin-M}
M_z^{S_z} = - \frac{\hbar}{2e} \sum_{\xi, \sigma = \pm } \xi \sigma
M_z (\Delta \to \Delta_{\xi \sigma}),
\end{equation}
which is nonzero even when the time-reversal symmetry is unbroken. Note that both the orbital magnetization
\begin{equation}
\label{M2-M}
M_z = \sum_{\xi, \sigma = \pm } \xi
M_z (\Delta \to \Delta_{\xi \sigma})
\end{equation}
and the electrical Hall conductivity
\begin{equation}
\label{sigma2Hall-Hall}
\sigma_{xy} = \sum_{\xi, \sigma = \pm } \xi
\sigma_{xy} (\Delta \to \Delta_{\xi \sigma})
\end{equation}
in silicene in the absence of a magnetic field are equal to zero.

The final thermospin coefficient $\beta_{xy}^{S_z}$ is given by
Eq.~(\ref{beta-clean}) with $\Delta$ replaced by $\Delta _{\text{SO}}$ and $\beta_0$ by $\beta_0^s = k_B/2$.
Obviously, everything stated above concerning a large Nernst signal for the model (\ref{Dirac-Hamiltonian})
turns out to be applicable for the SN effect in silicene.

\begin{figure}[ht]
\includegraphics[width=4.1cm]{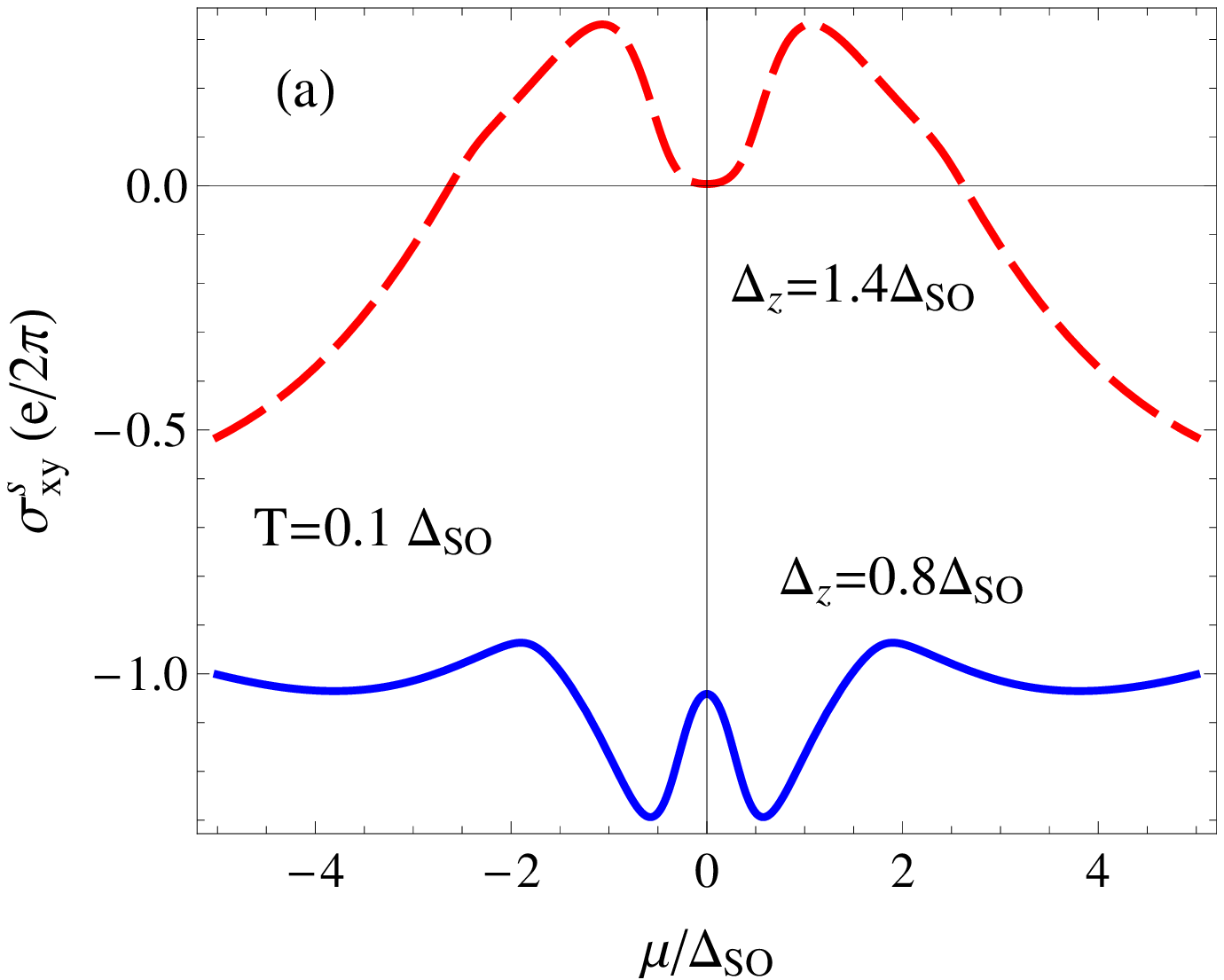}%\newline
\includegraphics[width=4.1cm]{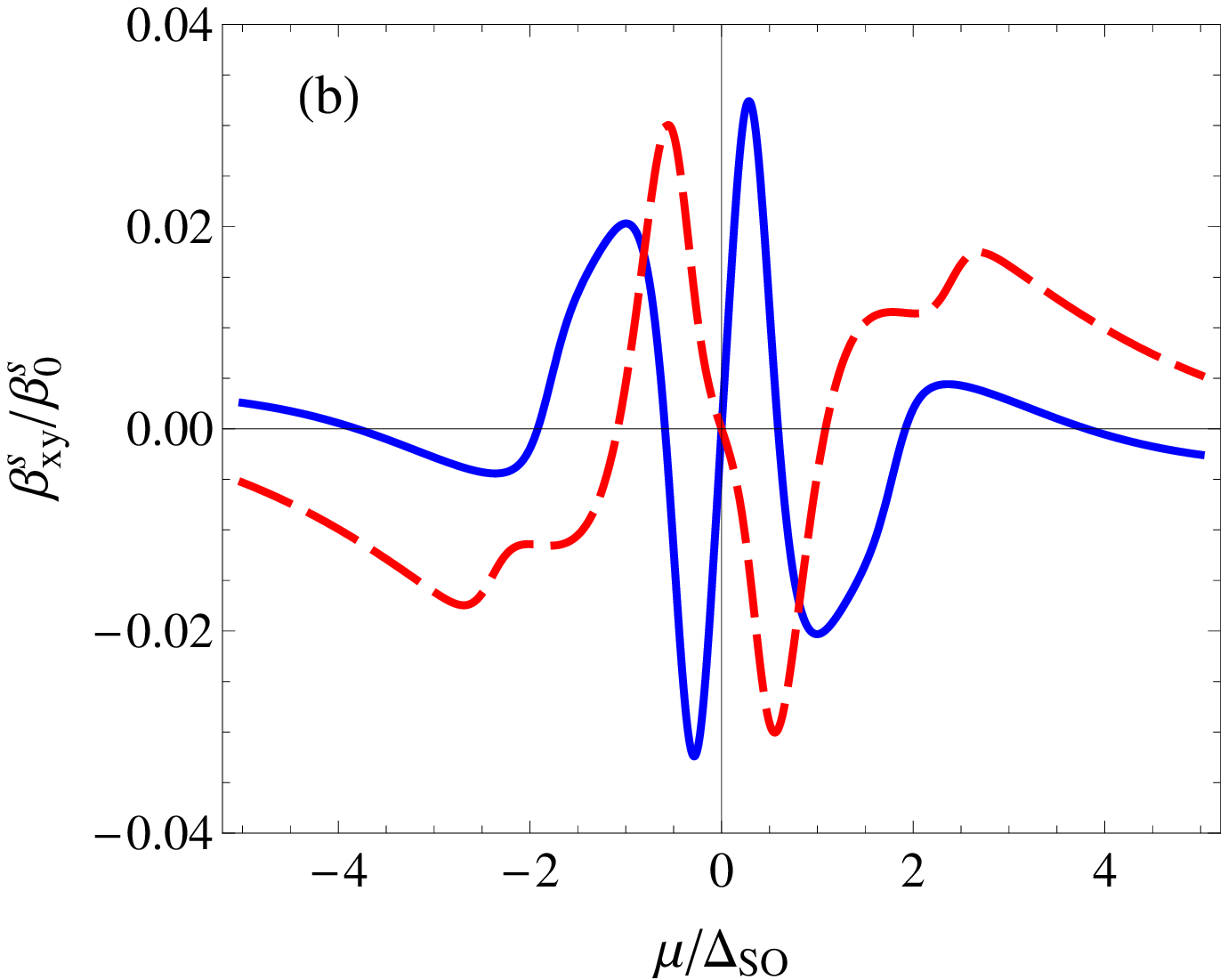}
\caption{(Color online) Left panel (a): spin Hall conductivity $\sigma_{xy}^{S_z}(\mu)$
in units of  $\sigma_0^s = e/(2 \pi)$;
right panel (b): thermospin coefficient $\beta_{xy}^{S_z}(\mu)$ in units of $\beta_0^s = k_B/2$
as functions of the chemical potential $\mu$ in the units of $\Delta _{\text{SO}} >0$.
%The two cases are shown: the solid (blue) curve is for
%and the dashed (red) curve is .
}
\label{fig:2}
\end{figure}

We present the dependences $\sigma_{xy}^{S_z}(\mu)$ and $\beta_{xy}^{S_z}(\mu)$
computed using the kernel (\ref{Hall-kernel-vertex}) for a  general case $\Delta_z \neq 0$
in the left and right panels of  Fig.~\ref{fig:2}.
The case with the sublattice asymmetry gap $\Delta_z = 0.8 \Delta _{\text{SO}}$ is shown by the solid (blue) curves
and the case $\Delta_z = 1.4 \Delta _{\text{SO}}$ is shown by the dashed (red) curves. We took the temperature
$T = 0.1 \Delta _{\text{SO}}$ that  corresponds to  $T \approx 10\, \mbox{K}$ for silicene.
We find that the presence of the disorder vertex resulted in a rather specific and rich pattern seen in the
$\beta_{xy}^{S_z}$ dependence, especially when the value of gap $\Delta_z$ is closer to  $\Delta _{\text{SO}}$.

\begin{figure}[h]
\centering{\includegraphics[width=8.5cm]{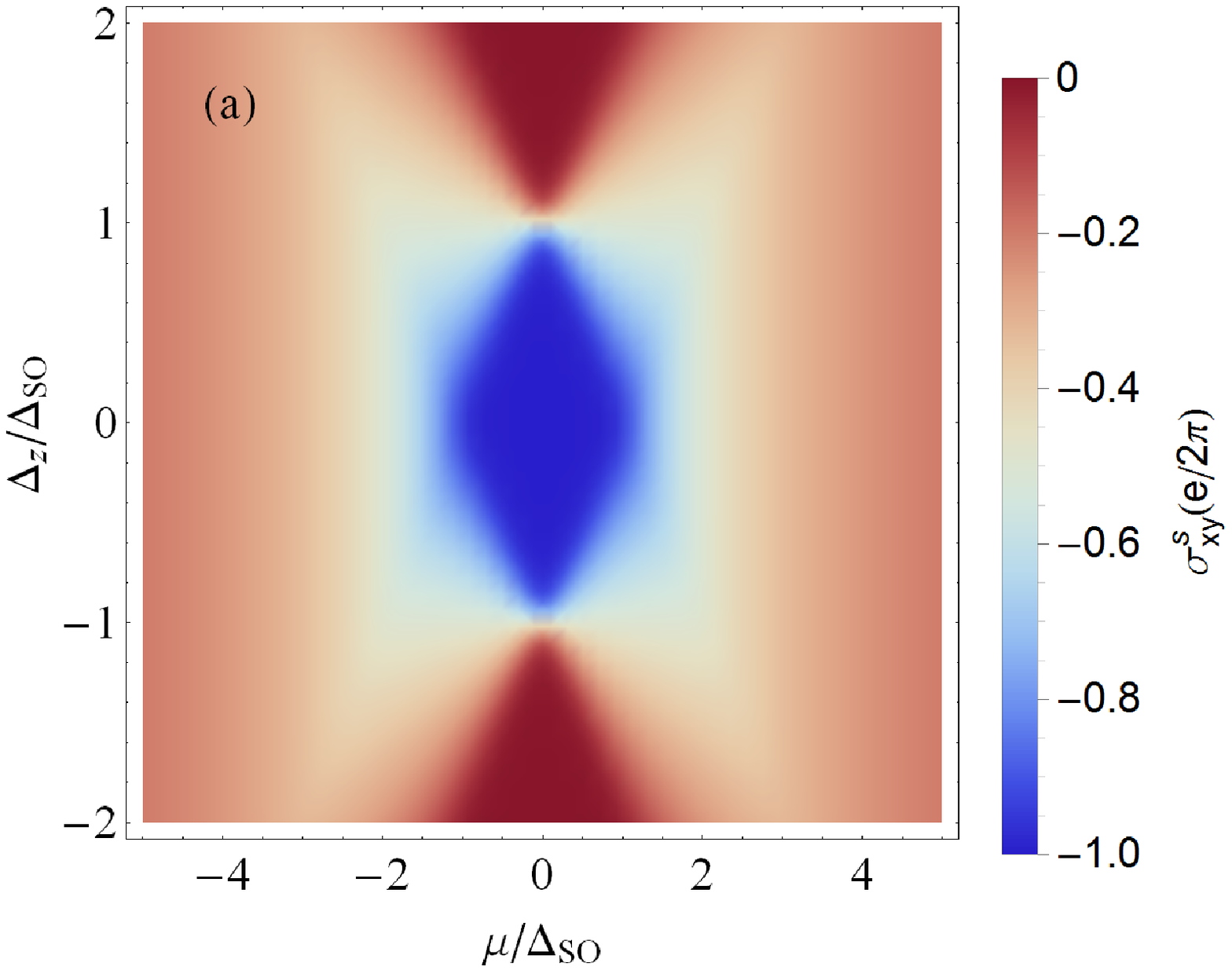}}\newline
\centering{\includegraphics[width=8.5cm]{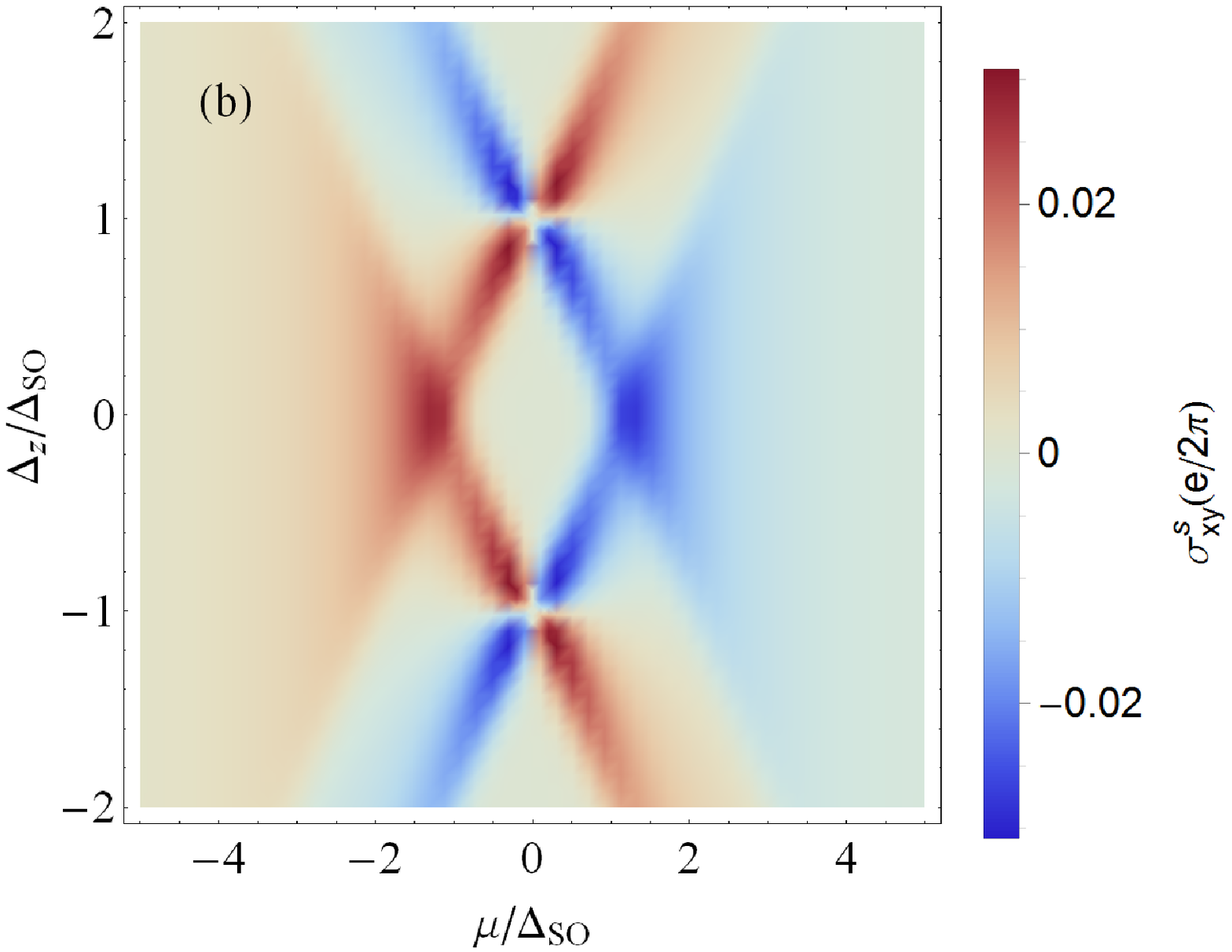}}
\caption{(Color online) Top panel: (a): spin Hall conductivity $\sigma_{xy}^{S_z}$ in units of the value $\sigma_0^s$;
bottom panel (b): thermospin coefficient $\beta_{xy}^{S_z}$ in units of the value $\beta_0^s$
as functions of the chemical potential  $\mu$ and the sublattice asymmetry gap $\Delta_z$ in the units of $\Delta _{\text{SO}}>0$
for the clean case.
}
\label{fig:3}
\end{figure}

\begin{figure}[h]
%\begin{center}
%\includegraphics[width=7.5cm]{fig3a.jpg}
%\includegraphics[width=7.5cm]{fig3b.jpg}
%\end{center}
\centering{\includegraphics[width=8.5cm]{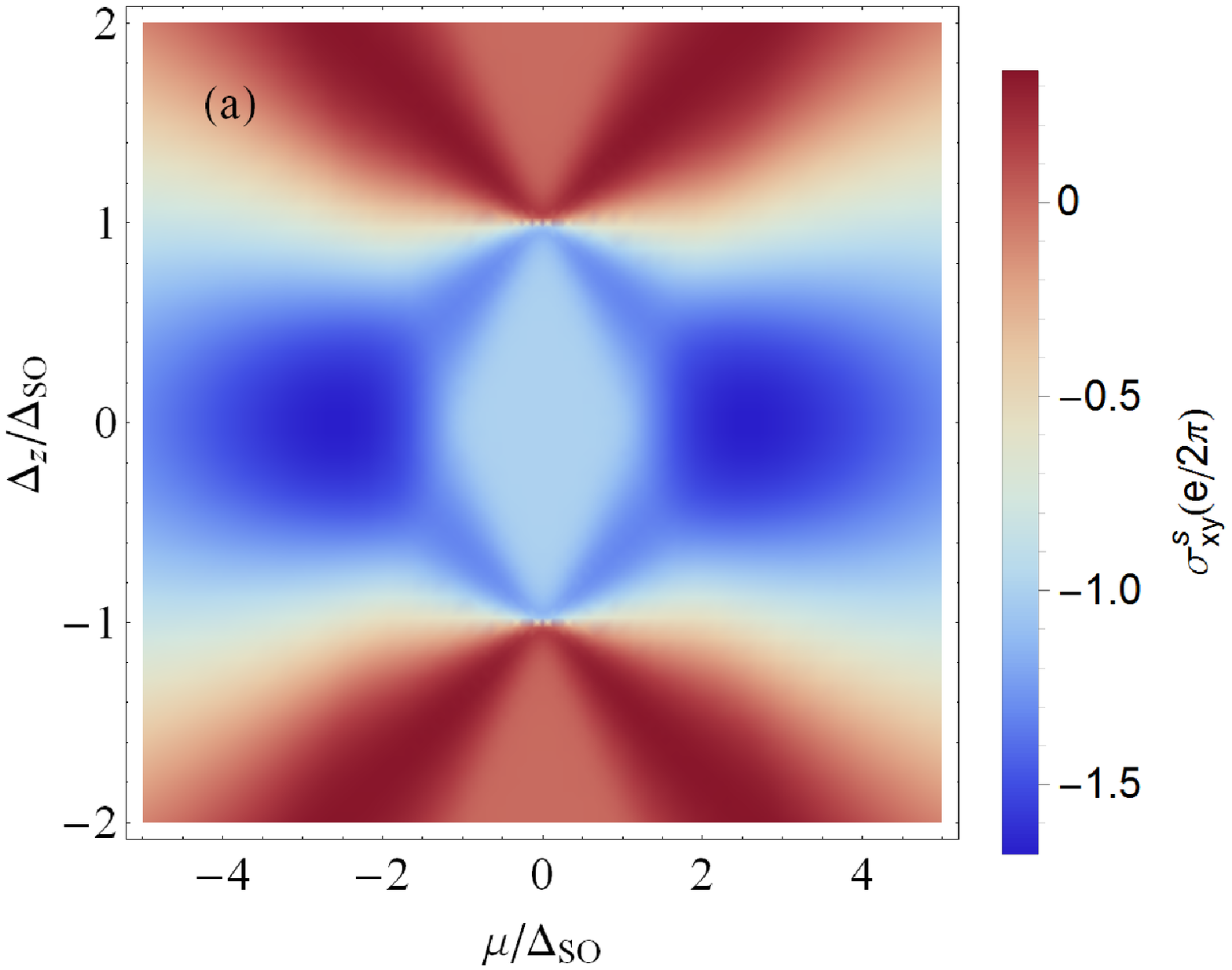}}\newline
\centering{\includegraphics[width=8.5cm]{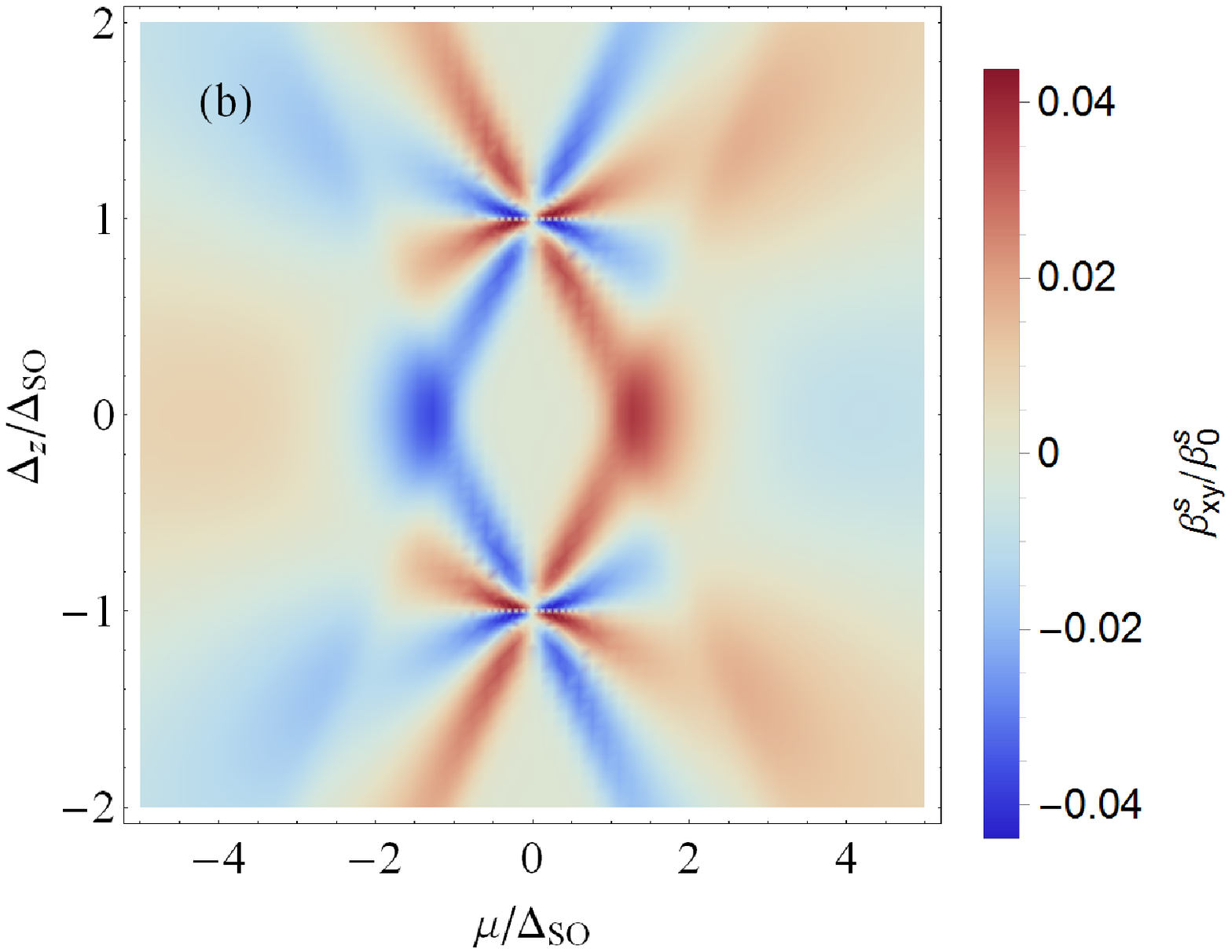}}
\caption{(Color online) Top panel: (a): spin Hall conductivity $\sigma_{xy}^{S_z}$ in units of $\sigma_0^s$;
bottom panel (b): thermospin coefficient $\beta_{xy}^{S_z}$ in units of  $\beta_0^s$
as functions of the chemical potential  $\mu$ and the sublattice asymmetry gap $\Delta_z$ in the units of $\Delta _{\text{SO}}>0$
for the case with disorder.
}
\label{fig:4}
\end{figure}

It is instructive to represent the dependences of $\sigma_{xy}^{S_z}$ and $\beta_{xy}^{S_z}$
as the function of both of their variables $\mu$ and $\Delta_z$
employing the density  plot.  Figures~\ref{fig:3} and \ref{fig:4}
are computed, respectively, using the kernel  $\mathcal{A}_{H}^{\left( cl\right) }$ for the clean case
and   the kernel $\mathcal{A}_{H}^{\left( d\right) }$ for the case with disorder.
One can see that in agreement with the analytical expressions for these kernels, the spin Hall conductivity
$\sigma_{xy}^{S_z}(\mu,\Delta_z)$ is even with respect to the variables $\mu$ and $\Delta_z$ both for the clean
and disordered cases. On the other hand, the thermospin coefficient $\beta_{xy}^{S_z}(\mu,\Delta_z)$ is odd with respect to
$\mu$ and even with respect to $\Delta_z$ in  both cases.
Note that the spin Hall conductivity computed in the clean limit
[Fig.~\ref{fig:3}(a)]  is very similar to that  found in Ref.~\onlinecite{Dyrdal2012PSS} (Fig.~2).
It is worthwhile to stress the crucial role of disorder that dramatically changes the character
of the dependences $\beta_{xy}^{S_z}(\mu,\Delta_z)$ and
$\sigma_{xy}^{S_z}(\mu,\Delta_z)$ shown in Fig.~\ref{fig:4}.
One can see that the number of the extrema near the points
$(\mu=0, \Delta_z = \pm \Delta _{\text{SO}})$  is duplicated with respect to the clean case. The same happens with the
thermospin coefficient.

Similarly to the case of  bilayer graphene, it should be possible in
the experiments on silicene also to vary both $\mu$ and $\Delta_z$ independently and observe the predicted structures.

\section{Conclusions}
\label{sec:disc}

In the present work we studied the thermospin effect in silicene which
is the base for other low-buckled Dirac materials such as germanene, stanene,
and phosphorene. Neglecting the Rashba term in the  Hamiltonian (\ref{Hamiltonian-density}),
we arrived at the simple but still nontrivial model of the two-component massive Dirac fermions.
So far such an approach has allowed us to make a lot of progress in the analytical studies of the anomalous Hall
effect \cite{Nagaosa2010RMP}. Our study showed that anomalous thermoelectric transport can also
be successfully investigated in this framework. We explicitly demonstrated how
the standard Kubo formula has to be altered by including the effective
magnetization leading to the correct off-diagonal thermoelectric coefficient. We also predicted
 a strong spin Nernst effect with nontrivial dependences on the carrier
concentration and electric field applied in silicene and other
low-buckled Dirac materials.

In conclusion, let us discuss the role of spin-nonconserving terms
omitted in the model Hamiltonian (\ref{Hamiltonian-density}).
Actually, there are two different terms \cite{Ezawa2012NJP}. The first of the
Rashba coupling contributions is associated with the nearest-neighbor
hopping term and is induced by the external electric field $E_z$.
The second term is associated with the next-nearest-neighbor hopping of electrons.

The first term in the continuum limit coincides with the Kane-Mele \cite{Kane2005PRL}
Rashba term. In the clean limit, its impact on the SN effect
is analyzed in detail in Ref.~\onlinecite{Bao2012ChinPhys} using the Mott formula.
In the existing literature on silicene, this term is neglected, \cite{Ezawa2012NJP} because
the corresponding coupling constant is two or three orders of magnitude less
than the value of the second Rashba term.

With regard to the second Rashba coupling contribution, its effect is suppressed by the fact
that it enters the dispersion law as the product of the coupling constant and momentum \cite{Ezawa2012NJP}.
The impact of this Rashba term on the spin Hall conductivity was considered in Ref.~\onlinecite{Dyrdal2012PSS}.
It is shown that in the reference case described by Eq.~(\ref{SH-ref}), the spin Hall conductivity is modified by the factor
$v_F^2/(v_F^2 + a^2 \lambda_{R2}^2/\hbar^2)$, where $\lambda_{R2}$ is the coupling constant and $a$ is the lattice constant.
Using that $v_F = \sqrt{3}/(2 \hbar) ta$ with $t \sim 1.6 \, \mbox{eV}$  being the nearest neighbor hopping parameter,
one can estimate that for the realistic values of  $\lambda_{R2} \sim 1 \, \mbox{meV}$ the impact of the Rashba term on the
spin Hall conductivity is indeed small, $\sim 10^{-6}$.
Moreover, since this term does not affect the dependence of the spin Hall conductivity on
$\mu$, the Mott formula would give the same result for the  SN effect.
Nevertheless, the role of the Rashba interactions, especially in the presence of the dressed by disorder vertex,
should be thoroughly studied.

The progress achieved in measuring spin currents (see e.g. Ref.~\onlinecite{Maekawa2013JPSJ} for a review)
allows us to hope that the predicted  very specific and strong SN effect in silicene can
also be observed.  At present theoretical studies of silicene and other related
Dirac materials are ahead of the experimental ones.
Silicene is only available on Ag and ZrB$_2$ \cite{Fleurence2012PRL} substrate which are both conductive,
there are no yet transport and optical measurements. Certainly the spintronics on silicene
will only be possible when a more conventional transport experiment is performed.

\begin{acknowledgments}

The authors acknowledge the support of the
European IRSES Grant SIMTECH No. 246937. The work
of V.P.G. and S.G.Sh. was also supported by the Science
and Technology Center in Ukraine (STCU) and the National
Academy of Sciences of Ukraine (NASU) within the framework
of the Targeted Research \& Development Initiatives
(TRDI) Program under Grant No. 5716-2 ``Development of
Graphene Technologies and Investigation of Graphene-based
Nanostructures for Nanoelectronics and Optoelectronics.''

\end{acknowledgments}

\appendix

\section{Calculation of the magnetization}
\label{sec:Appendix-magnetization}

The equation for the GF
in the external magnetic field $B$ has the form
\begin{equation}
\begin{split}
\label{G-eq}
& \int d\mathbf{r}'\left[\omega-\pmb{\tau}\left(-i\hbar\pmb{\nabla} +\frac{e}{c}\mathbf{A}(\mathbf{r})\right)
-\Delta\tau_3\right]_r\delta(\mathbf{r}-\mathbf{r}')\\
& \times G(\omega;\mathbf{r}',\mathbf{r}'')
=\delta(\mathbf{r}-\mathbf{r}''),
\end{split}
\end{equation}
where we use the symmetric gauge $A_i=-(B/2)\epsilon_{ij}r_j$,
with  $\epsilon_{ij}$ being the antisymmetric tensor. We set  $\mu=0$ for  brevity.
The GF $G(\omega;\mathbf{r}',\mathbf{r}'')$ can be rewritten in the form
\begin{equation}
G(\omega;\mathbf{r}',\mathbf{r}'')=e^{i\Phi(\mathbf{r'},\mathbf{r}'')}\tilde{G}(\omega;\mathbf{r}'-
\mathbf{r}''),
\end{equation}
where $\tilde{G}(\omega;\mathbf{r}'-
\mathbf{r}'')$ is the translation invariant
part and
\begin{equation}
\Phi(\mathbf{r'},\mathbf{r}'')=\frac{e}{\hbar c}
\int\limits_{\mathbf{r}'}^{\mathbf{r}''}d\mathbf{z}\mathbf{A}(\mathbf{z})=-\frac{e}{\hbar c}\mathbf{r'}\mathbf{A}(\mathbf{r}'')
\end{equation}
is the Schwinger phase.
Inserting the phase factor along with the $\delta$ functions we rewrite Eq.~(\ref{G-eq})
in the form
\begin{widetext}
\begin{equation}
\begin{split}
\int d\mathbf{r}'\left[\omega-\pmb{\tau}\left(-i\hbar\pmb{\nabla}_r +\frac{e}{c}\mathbf{A}(\mathbf{r})\right)
-\Delta\tau_3\right]
\left[e^{-i \frac{e}{\hbar c}\mathbf{r}\mathbf{A}(\mathbf{r}')}
\delta(\mathbf{r}-\mathbf{r}')\right]e^{-i \frac{e}{\hbar c}\mathbf{r'}\mathbf{A}(\mathbf{r}'')}
\tilde{G}(\omega;\mathbf{r}'-\mathbf{r}'')
=\delta(\mathbf{r}-\mathbf{r}'')e^{-i \frac{e}{\hbar c}\mathbf{r}\mathbf{A}(\mathbf{r}'')}.
\end{split}
\end{equation}
Taking the exponential factor to the left-hand side, the three phases combine together,
which gives the magnetic flux threading through the triangle formed by the three points $\mathbf{r}'$,
$\mathbf{r}'$,  and $\mathbf{r}''$,
\begin{equation}
\begin{split}
\int d\mathbf{r}'\left[\omega-H(\mathbf{r}-\mathbf{r}')\right]
\delta(\mathbf{r}-\mathbf{r}')
\left[e^{-i \frac{e}{\hbar c}(\mathbf{r}-\mathbf{r}')\mathbf{A}(\mathbf{r}'-\mathbf{r}'')}
\tilde{G}(\omega;\mathbf{r}'-\mathbf{r}'')\right]=\delta(\mathbf{r}-\mathbf{r}''),
\end{split}
\label{eq:GF_translation}
\end{equation}
where we introduced
\begin{equation}
H(\mathbf{r}-\mathbf{r}') =
\pmb{\tau}\left(-i\hbar\pmb{\nabla}_{r-r'} +\frac{e}{c}
\mathbf{A}(\mathbf{r}-\mathbf{r}')\right) +\Delta\tau_3
\end{equation}
and used the relationship
\begin{equation}
\mathbf{x}\mathbf{A}(\mathbf{z})+\mathbf{z}\mathbf{A}(\mathbf{y})-\mathbf{x}\mathbf{A}(\mathbf{y})
=(\mathbf{x}-\mathbf{z})\mathbf{A}(\mathbf{z}-\mathbf{y}).
\end{equation}
Eq.~(\ref{eq:GF_translation}) is now translation invariant, and we solve it with respect to $\tilde{G}$
by expanding the exponent to the first order in $B$,
\begin{equation}
\label{GF-expanded}
\begin{split}
&\int d\mathbf{r}'\left[\omega-H(\mathbf{r}-\mathbf{r}')\right]\delta(\mathbf{r}-\mathbf{r}')
\tilde{G}(\omega;\mathbf{r}'-\mathbf{r}'')
 -\frac{ie}{\hbar c}\int d\mathbf{r}'\left(\left[\omega-H(\mathbf{r}-\mathbf{r}')\right]\delta(\mathbf{r}-\mathbf{r}')\right)
(\mathbf{r}-\mathbf{r}')\mathbf{A}(\mathbf{r}'-\mathbf{r}'')\tilde{G}(\omega;\mathbf{r}'-\mathbf{r}'') \\
&=\delta(\mathbf{r}-\mathbf{r}'').
\end{split}
\end{equation}
Since both terms on the left-hand side of Eq.~(\ref{GF-expanded}) have a form of the convolution,
it can be solved using the Fourier transform
\begin{eqnarray}
F.T.[(\omega-H(\mathbf{r}))\delta(\mathbf{r})](\mathbf{k})\tilde{G}(\omega;\mathbf{k})+
\frac{i e B}{2\hbar c}\epsilon_{ij}F.T.[r_i(\omega-H(\mathbf{r}))\delta(\mathbf{r})](\mathbf{k})
F.T.\left(r_j\tilde{G}(\omega;\mathbf{r})\right)(\mathbf{k})=1,
\end{eqnarray}
\end{widetext}
where the F.T. is defined by
\begin{equation}
F.T.[f(\mathbf{r})](\mathbf{k})=\int\,d\mathbf{r}f(\mathbf{r})e^{-i\mathbf{k}\mathbf{r}}.
\end{equation}
Since the coordinate $r_i$ is replaced by the derivative $i\partial/\partial k_i$ we obtain
\begin{equation}
(\omega-\mathcal{H} (\mathbf{k}))\tilde{G}(\omega;\mathbf{k})-\frac{i e B}{2\hbar c}\epsilon_{ij}
\frac{\partial(\omega- \mathcal{H}(\mathbf{k}))}{\partial k_i}\frac{\partial\tilde{G}(\omega;\mathbf{k})}
{\partial k_j}=1,
\label{eq:tildeG}
\end{equation}
where the Hamiltonian density $\mathcal{H} (\mathbf{k})$  is given by Eq.~(3), so that we restored the
chemical potential $\mu$. Solving Eq.~(\ref{eq:tildeG}) to the first order in $B$, one obtains
\begin{equation}
\label{GF-H-final}
\tilde{G}=G_0 + \frac{i e B}{2\hbar c}\epsilon_{ij}G_0\frac{\partial G_0^{-1}}{\partial k_i}
\frac{\partial G_0}{\partial k_j},
\end{equation}
where $G_0 \equiv G_0(\omega, \mathbf{k})=[\omega- \mathcal{H}(\mathbf{k})]^{-1}$.
The GF (\ref{GF-H-final}) can be compared with Eq.~(7) of  Ref.~\onlinecite{Chen2011PRB}.

Starting from this expression, it is
straightforward to calculate the number density at the finite temperature in the magnetic field:
\begin{widetext}
\begin{equation}
%\begin{split}
\rho(\mu,T,H)= \rho_0(\mu,T) -
\frac{i e B}{2\hbar c}\epsilon_{ij}
T\sum\limits_{\epsilon}\int\frac{d^2k}{(2\pi)^2}{\rm tr}\left(G_0\frac{\partial G_0^{-1}}
{\partial k_i}G_0\frac{\partial G_0^{-1}}{\partial k_j}G_0\right),
%\end{split}
\label{rho-linearB}
\end{equation}
where now the GF is written in the Matsubara representation $G_0 = G_0(\epsilon_n, \mathbf{k})$.
The magnetization $M_z = - \partial \Omega/\partial B$ is obtained from the
thermodynamic potential $\Omega(\mu,T,B)$ which in its turn is derived
by integrating the relationship $\rho = -\partial \Omega/\partial \mu $ over $\mu$.
The first term in Eq.~(\ref{rho-linearB}) does not contribute to the magnetization,
therefore, we consider only the second term that can be rewritten in the form
\begin{equation}
\label{density-general}
%\begin{split}
 \rho_M(\mu,T,B)=-\frac{i e\hbar v_F^2 B}{2 c}\epsilon_{ij}
  T \sum_{\epsilon_n} \int\frac{d^2k}{(2\pi)^2}{\rm tr}\left[G_0(\epsilon_n,\mathbf{k})\tau_i
G_0(\epsilon_n,\mathbf{k})\tau_j G_0(i\epsilon_n,\mathbf{k})\right].
%\end{split}
\end{equation}
We derived the following rather simple expression for
the carrier density
\begin{equation}
%\begin{split}
 \rho_M(\mu,T,B)=\frac{e B\Delta}{4\pi^2\hbar c|\Delta|}
 {\rm Im}\left[\Psi\left(\frac{1}{2}+
\frac{\Gamma_0-i(\mu-|\Delta|)}{2\pi T}\right)-\Psi\left(\frac{1}{2}+
\frac{\Gamma_0-i(\mu+|\Delta|)}{2\pi T}\right)\right],
%\end{split}
\end{equation}
where $\Psi(z)$ is  the digamma function and
we took into account the effect of level broadening caused by impurities
$\Sigma ^{R}(\epsilon )=-i\Gamma _{0}(\epsilon )\tau _{0}$ with $\Gamma _{0}(\epsilon ) = \Gamma_0$.
Integrating over $\mu$ and differentiating over $B$ we arrive at the final result
\begin{equation}
\label{M-Gamma}
\begin{split}
M_z = \frac{eT \,{\rm sgn} \,(\Delta)}{2\pi\hbar c}{\rm Re}& \left[\ln\Gamma\left(\frac{1}{2}+
\frac{\Gamma_0-i(\mu-|\Delta|)}{2\pi T}\right)-\ln\Gamma\left(\frac{1}{2}+
\frac{\Gamma_0-i(\mu+|\Delta|)}{2\pi T}\right)\right.\\
& -\left.\ln\Gamma\left(\frac{1}{2}+
\frac{\Gamma_0+i|\Delta|}{2\pi T}\right)+\ln\Gamma\left(\frac{1}{2}+
\frac{\Gamma_0-i|\Delta|}{2\pi T}\right)\right],
\end{split}
\end{equation}
\end{widetext}
$\Gamma(z)$ is  the gamma function. For $\Gamma_0 =0$ using the relationship
\begin{equation}
\Gamma\left(\frac{1}{2}+ix\right)\Gamma\left(\frac{1}{2}-ix\right)=\frac{\pi}{\cosh(\pi x)},
\end{equation}
we obtain Eq.~(\ref{M-final}) of the main text.

Note that Eq.~(\ref{M-final})  for the orbital magnetization can be rewritten in the form
of the general expression for the magnetization that was suggested in the studies of the role of the Berry phase
in the anomalous thermoelectric transport
(see, for example, Eq.~(6) and (14) in Refs.~\onlinecite{Xiao2006PRL} and \onlinecite{Shi2007PRL}, respectively,
and Refs.~\onlinecite{Xiao2010RMP,Nourafkan2014})
\begin{equation}
\begin{split}
\mathbf{M}=& \sum_{n}\int \frac{d\mathbf{k}}{(2 \pi)^2}\bigg[ \mathbf{m}_n(\mathbf{k}) f (\epsilon_n(\mathbf{k})-\mu)  \\
& \left.
+\frac{e}{\hbar}\pmb{\Omega}_n(\mathbf{k})\frac{1}{\beta}\ln\left(1+
e^{-\beta(\epsilon_n(\mathbf{k})-\mu)}\right)\right].
\end{split}
\label{magnetization-general}
\end{equation}
Here
\begin{equation}
\mathbf{m}_n(\mathbf{k}) = - \frac{i e}{2\hbar c} \langle {\pmb\nabla}_k u_{nk}|[\mathcal{H} (\mathbf{k})-\epsilon_n(\mathbf{k})]
|\pmb{\nabla}_k u_{nk}\rangle
\end{equation}
is the orbital magnetic moment of the state $(n,\mathbf{k})$,
$\pmb{\Omega}_n(\mathbf{k})=i\langle \pmb{\nabla}_k u_{nk}|\times| \pmb{\nabla}_k u_{nk}\rangle$
is the Berry curvature, $n$ is the band index, $|u_{nk}\rangle$ is the band wave function,
and $\beta = 1/(k_B T)$.

In the considered case $n=\pm$, $\epsilon_{\pm}(\mathbf{k})=\pm a$,
and the wave function is
\begin{equation}
|u_{\pm k}\rangle=\frac{1}{\sqrt{2a(a\mp\Delta)}}\left(\begin{array}{c}\hbar v_F(k_x-ik_y)\\
\pm a-\Delta\end{array}\right).
\end{equation}

Going back to Eq.~(\ref{M-final}) one can see that at $T =0$ it reduces to the expression
\begin{equation}
M_z=\frac{e}{4\pi\hbar c}\left[\Delta \, {\rm sgn} \, (\mu) \theta(|\mu|-|\Delta|)
+\mu \, {\rm sgn} \, (\Delta) \theta(|\Delta|-|\mu|)\right]
\end{equation}
so that $c M_z/T$ exactly cancels out the diverging part ${\tilde \beta}_{xy}$ given by Eq.~(\ref{tilde-beta}).

\end{document}